\documentclass{elsart}

\usepackage{amssymb}
\usepackage{graphicx}
\usepackage{epsfig}
\newcommand{\ta}[1]{#1\hspace{-.42em}/\hspace{-.07em}} 
\newcommand{\tb}[1]{#1\hspace{-.93em}/\hspace{.39em}} 
\newcommand{\taa}[1]{#1\hspace{-.55em}/\hspace{-.09em}} 
\newcommand{\beq}{\begin{equation}} 
\newcommand{\eeq}{\end{equation}}   
\newcommand{\bea}{\begin{eqnarray}} 
\newcommand{\eea}{\end{eqnarray}}
\newcommand{\non}{\nonumber}

\begin{document}

\begin{frontmatter}



\title{ The reaction 
 {\mathversion{bold}
  {${ \mathbf{ e^+e^-\to e^+ e^-\pi^+\pi^-}}$}}
 and the pion form factor measurements via the radiative return method
 \thanksref{labelf}}
 \thanks[labelf]{Work supported in part by 
 EC 5-th Framework EURIDICE network  project HPRN-CT2002-00311,
 TARI project RII3-CT-2004-506078 and 
 Polish State Committee for Scientific Research (KBN)
under contract 1 P03B 003 28. }


\author{Henryk Czy\.z,}
\author{El\.zbieta Nowak-Kubat}

\address{Institute of Physics, University of Silesia,
PL-40007 Katowice, Poland.}

\begin{abstract}
 The role of the reaction $e^+e^-\to e^+ e^-\pi^+\pi^-$ in the pion
 form factor measurements via radiative return method without photon
 tagging is studied in detail. The analysis is based on the developed
 Monte Carlo program EKHARA, which ingredients are also presented.
\end{abstract}

\end{frontmatter}

\section{Introduction}
\label{intro}

 Radiative return method of the hadronic cross section extraction
 from the measurement of the cross section of the reaction
 $e^+e^-\to {\rm hadrons}+{\rm photon(s)}$, proposed already some time
 ago \cite{Zerwas}, is currently being used by KLOE \cite{KLOE1}
 and BaBar \cite{BABAR} providing very precise experimental data.
 Further improvement in accuracy is crucial
 for predictions of the hadronic contributions to $a_\mu$, the anomalous 
 magnetic moment of the muon, as the error on the hadronic
 contributions to $a_\mu$ may obscure possible new physics signal,
 seen as a deviation from the Standard Model (SM) predictions.
 The same information is essential for the evaluation of
 the running of the electromagnetic
 coupling ($\alpha_{QED}$) from its value at low energy up
 to $M_Z$ as the present error on the hadronic contributions
 is too big to fully profit from the data of the future ILC
 (International Linear Collider) running in the gigaZ mode.
 For recent reviews  of these subjects look
 \cite{Davier:2003pw,Jegerlehner:2003rx,Nyffeler:2004mw}.
 
 The extraction of the cross section $\sigma(e^+e^-\to {\rm hadrons})$
 from the measured cross section 
 $\sigma(e^+e^-\to {\rm hadrons} + {\rm photons})$ relies on the
 factorization
\bea
{ d\sigma(e^+e^- \to \mathrm{hadrons} + n \gamma) = }
{\ H }{ \ d\sigma(e^+e^-\to \mathrm{hadrons})} \ ,
\label{radiator}
\eea
valid at any order for photons emitted from initial leptons, where
 the function $H$ contains QED radiative corrections. This function is known
 analytically, if no cuts are imposed, at next to leading order
 (NLO) and has to be provided
 in form of an event generator \cite{PHOKHARA,Czyz:2005as} of the reaction  
 $e^+e^-\to {\rm hadrons} + {\rm photons}$ 
 for a realistic experimental setup.

 Let us focus on the most important process, where the 'hadrons'
 means just $\pi^+\pi^-$ pair. This reaction gives the dominant contribution
 to the hadronic part of $a_\mu$ as well as to its error.
 In case the photon(s) are not measured and only charged pions are tagged,
 there exists a number of possible backgrounds. It was pointed out
 in \cite{Hoefer:2001mx}, basing on integrated over the whole phase space
 analytical
 formulae and 
 containing contributions from diagrams (a) and (d) in Fig.\ref{diag},
  that the reaction $e^+e^-\to e^+ e^-\pi^+\pi^-$ 
 can give sizable contributions to the radiative return process, 
 especially for low invariant masses of two pion system. To examine
 this contribution for a realistic experimental setup, a Monte Carlo
 program EKHARA was developed. It is 
 the scope of this letter to present its ingredients and results
 of the simulations
 relevant for the pion form factor extraction via the radiative return method.

 Some partial results concerning the electron--positron pair production
 contributions to the pion form factor measurements  and tests of the code
 were presented in \cite{n1,n2}, while in this letter the amplitude 
 describing the reaction $e^+e^-\to e^+ e^-\pi^+\pi^-$ is discussed,
 generation procedure is described in detail and results based on the
 complete tree level amplitude are presented.

\section{The scattering amplitude and the generation procedure}

\begin{figure}[ht]
\begin{center}
\epsfig{file=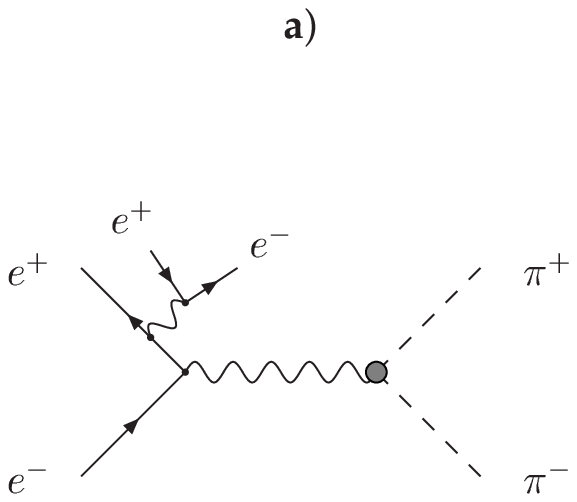,width=3.3cm,height=3.2cm}
\hskip+0.5cm
\epsfig{file=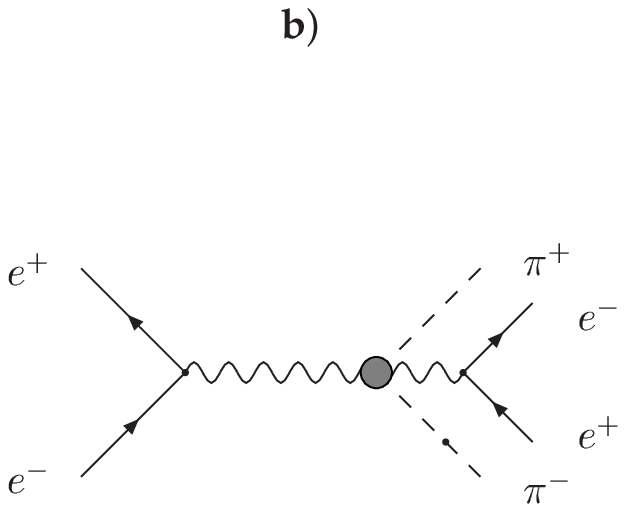,width=3.2cm,height=3.2cm}
\hskip+0.2cm
\epsfig{file=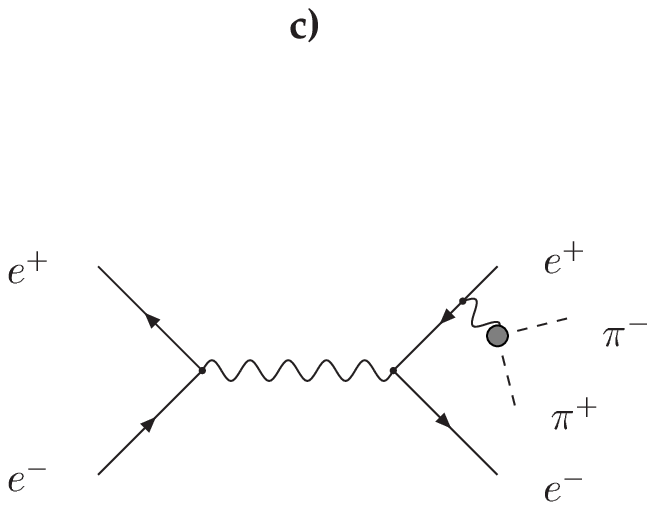,width=3.3cm,height=3.2cm}
\hskip+2cm
\epsfig{file=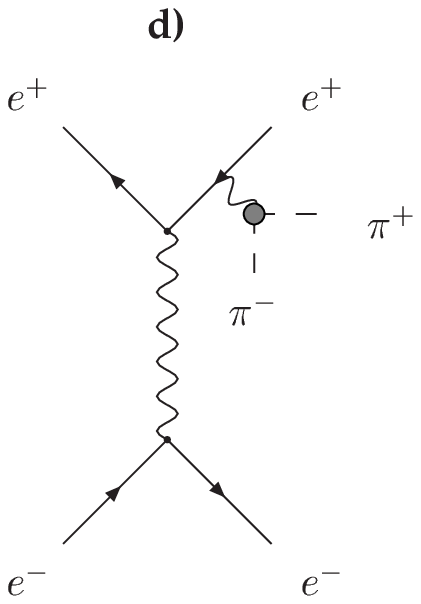,width=2.2cm,height=3.2cm}
\hskip+0.5cm
\epsfig{file=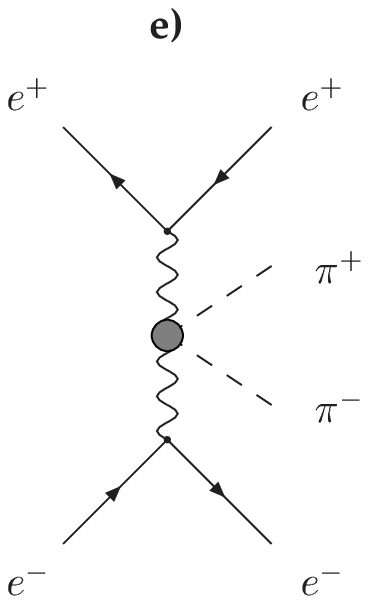,width=2cm,height=3.2cm}
\end{center}
\caption{Diagrams contributing to the process 
$e^+(p_1)e^-(p_2) \to \pi^+(\pi_1)\pi^-(\pi_2)e^+(q_1)e^-(q_2)$:
 initial state pair emission (a),
 final state pair emission (b,c),
 pion pair emission from t--channel Bhabha process  (d) and
 $\gamma^*\gamma^*$ pion pair production (e).
 Only one representative diagram for a given set of diagrams is shown. 
}
\label{diag}
\end{figure}

 As stated already in the introduction, the reaction 
  $e^+e^-\to e^+ e^-\pi^+\pi^-$ plays a role in the pion form factor
 measurement via the radiative return method only if the photons
 in the reaction $e^+e^-\to \pi^+ \pi^- \ + \ {\rm photons}$ are not
 tagged. This version of the radiative return method was used already
 by KLOE \cite{KLOE1} and as
 more accurate analysis, based on a significantly bigger
  data sample, is expected, a detailed study of all possible contributions
 is obligatory. The complete set of the lowest order diagrams,
 describing the reaction  $e^+e^-\to e^+ e^-\pi^+\pi^-$, is
 shown schematically in Fig.\ref{diag}. 

 Helicity amplitude method, with the conventions described
 in \cite{Kolodziej:1991pk,Jegerlehner:2000wu},
 was used for the scattering amplitude  
 evaluation. It allows for a fast numerical evaluation and, in addition, all
 interferences are easily included.
 Moreover, it partly
 avoids numerical cancellations present, when one uses the trace
 method to get the square of the amplitude. To model photon--pion
 interactions, we use scalar QED (sQED) combined with the 
 vector dominance model (VDM). Within these assumptions,
 the amplitude has the form
\bea
 M = M_a + M_b+M_c+M_d+M_e \ ,
\eea 
where the amplitudes $M_i,\ \ i=a,\cdots, e$ correspond to the contributions
 from the diagrams (a)-(e) from Fig.\ref{diag}. They read

\bea
\kern-24pt M_a =
-\frac{ie^4}{k_1^2 Q^2}
\bar{u}(q_2)\gamma_{\mu}v(q_1) \cdot
\bar{v}(p_1)\left( 
 \frac{ (\gamma^{\mu}\not{k_1}-2p_1^{\mu})\taa{\Gamma}}
{k_1^2-2k_1\cdot p_1} 
+ \frac{\taa{\Gamma}(2p_2^{\mu}-\not{k_1}\gamma^{\mu})}
{k_1^2-2k_1\cdot p_2} 
\right)u(p_2)
\eea
\bea
 M_b=&&\frac{-2ie^4 F(s)}{s k_1^2 } 
\bar{v}(p_1)\gamma_{\mu} u(p_2) \\
&&\cdot \bar{u}(q_2)\left( \gamma^{\mu} 
+\frac{2\pi_{2}^\mu{\ta \pi_1}}{k_1^2+2\pi_1\cdot k_1}
+\frac{2\pi_{1}^\mu{\ta \pi_2}}{k_1^2+2\pi_2\cdot k_1}
\right)v(q_1) \non
\eea

\bea
\kern-24pt M_c=\frac{ie^4}{s Q^2}
\bar{u}(q_2)\left( 
 \frac{\gamma^{\mu}(\taa{\Gamma}\taa{Q}-2q_1\cdot \Gamma )}
{Q^2+2Q\cdot q_1}
+\frac{(\taa{\Gamma}\taa{Q}+2q_2\cdot \Gamma )\gamma^{\mu}}
{Q^2+2Q\cdot q_2}
\right)v(q_1) \cdot
\bar{v}(p_1)\gamma_{\mu} u(p_2)
\eea

\bea
\kern-24ptM_d= &
\frac{ie^4}{t Q^2} \bar{v}(p_1)\left(
 \frac{(\taa{\Gamma}\taa{Q}-2p_1\cdot \Gamma )\gamma^{\mu}}
{Q^2-2Q\cdot p_1}
+\frac{\gamma^{\mu}(\taa{\Gamma}\taa{Q}-2q_1\cdot \Gamma )}
{Q^2+2Q\cdot q_1}
\right)v(q_1) \cdot \bar{u}(q_2)\gamma_{\mu} u(p_2)\nonumber \\  &
-\frac{ie^4}{t_1 Q^2} \bar{u}(q_2)\left(
 \frac{\gamma^{\mu}(2p_2\cdot \Gamma -\taa{Q}\taa{\Gamma})}
{Q^2-2Q\cdot p_2}
+\frac{(2q_2\cdot \Gamma +\taa{\Gamma}\taa{Q})\gamma^{\mu}}
{Q^2+2Q\cdot q_2}
\right)u(p_2) \bar{v}(p_1)\gamma_{\mu} v(q_1)  
\eea

\begin{eqnarray}
M_e=&\frac{-2ie^4 F(t)F(t_1)}{t t_1 }\Biggl( 
\bar{v}(p_1)\gamma^{\mu} v(q_1) \bar{u}(q_2)\gamma_{\mu} u(p_2) \\ \nonumber &
+\frac{2\bar{v}(p_1)\tb{\pi_1}v(q_1) \bar{u}(q_2)\tb{\pi_2}u(p_2)}
 {t_1-2\pi_1(p_1-q_1)}
+\frac{2\bar{v}(p_1)\tb{\pi_2}v(q_1) \bar{u}(q_2)\tb{\pi_1}u(p_2)}
{t+2\pi_1(q_2-p_2)}
\Biggr) \ , \nonumber
\end{eqnarray}
where
\begin{equation}
\Gamma^{\mu} =  F(Q^2)(\pi_1-\pi_2)^{\mu}\ , \ k_1=q_1+q_2\ ,
 \ Q=\pi_1+\pi_2\ , 
\end{equation}
\begin{equation}
s=(p_1+p_2)^2\ , \ t=(q_2- p_2)^2 \ , t_1=(p_1- q_1)^2
\end{equation}
and $ F(Q^2)$ is the pion form factor, which was adopted from \cite{Khodj04}.
For the numerical evaluation, following again the method
 described in papers \cite{Kolodziej:1991pk,Jegerlehner:2000wu},
 the amplitude is rewritten into
 a form where only 2x2 matrices appear.

 We use the 
multi--channel variance reduction method to improve efficiency of the
 generator and the generation is split into four channels, where two of them
 absorb peaks present in t--channel diagrams and other two take care
 of the s--channel peaks.

 For the t--channel peaks absorption, we use the following
 phase space representation  

\bea
&&\kern-20pt{\rm dLips}_4(p_1+p_2;q_1,q_2,\pi_1,\pi_2) = \non \\
 &&{\rm dLips}_2(p_1+p_2;Q',q_2)
     \frac{dQ'^2}{2\pi}{\rm dLips}_2(Q';Q,q_1)
        \frac{dQ^2}{2\pi}{\rm dLips}_2(Q;\pi_1,\pi_2)
\eea

in one of the channels and an analogous one, with $q_1\leftrightarrow q_2$,
 in the other channel. As both channels are completely symmetric under
  $q_1\leftrightarrow q_2$, we will describe here only changes of variables,
 which smoothen the distribution, only in one of them.
 For the two invariant masses ($Q^2$ and $Q'^2$) the
  following change of variables was performed 
\bea
\kern-20ptQ^2 =  \left(\sqrt{s}-2m_e-\sqrt{-2z}\right)^2, \ \ 
z = -\frac{1}{2}(\sqrt{s}-2m_e-2m_\pi)^2(1-r_{Q^2}) \ ,
\eea
 \bea
\kern-20ptQ'^2 &=& \frac{1}{^3\sqrt{-3y}}+m_e^2\ ,  \\
\kern-20pty &=&-\frac{1}{3(Q^2+2\sqrt{Q^2}m_e)^3}
 +\left(\frac{1}{3(Q^2+2\sqrt{Q^2}m_e)^3}
 -\frac{1}{3(s-2\sqrt{s}m_e)^3}\right)r_{Q'^2} \ . \non
\eea

 The angles of $\bar q_2$ vector are defined in the initial $e^+e^-$ center
 of mass (cms) frame with z-axis along $\bar p_1$ 
 and the polar angle is used to absorb the peak coming
 from the propagator of the photon exchanged in the $t$-channel
\bea
\cos\theta_{q_2} &=& \frac{3m_e^2-s+Q'^2-2t}
{\sqrt{1-\frac{4m_e^2}{s}}\lambda^{1/2}(s,Q'^2,m_e^2)} \ , \
t = -\frac{1}{y}\ , \non \\
y &=& -\frac{1}{t_{-}} +\frac{s\sqrt{1-\frac{4m_e^2}{s}}
\lambda^{1/2}(s,Q'^2,m_e^2)}{m_e^2(Q'^2-m_e^2)^2}r_{\theta_{q_2}} \ , \
\phi_{q_2} = 2\pi r_{\phi_{q_2}} \ ,
\eea
where
\begin{equation}
t_{-} = \frac{1}{2}\left(3m_e^2-s+Q'^2-\sqrt{1-\frac{4m_e^2}{s}}
\lambda^{1/2}(s,Q'^2,m_e^2)\right) \ .
\end{equation}

The angles of the $\bar Q$ vector are defined 
in $Q'$ rest frame and the appropriate change of variables reads 

\bea
\cos\theta_Q &=& \frac{2E'_{1}Q_0-Q^2-\frac{1}{2|\bar{p'_1}||\bar{Q}|x}}
{2|\bar{p'_1}||\bar{Q}|}\ , \ \phi_{Q} = 2\pi r_{\phi_{Q}}  \\
 &&\kern-50pt x = \frac{-1}{2|\bar{p'_1}||\bar{Q}|(Q^2-2E'_{1}Q_0-2|\bar{p'_1}||\bar{Q}|)}
+ \frac{2}{(Q^2-2E'_{1}Q_0)^2-4|\bar{p'_1}|^2|\bar{Q}|^2}r_{\theta_Q}
 \ , \non
\eea

where $ E'_{1}= \frac{Q'^2-t+m_e^2}{2\sqrt{Q'^2}}$,
$Q_0 = \frac{Q'^2+Q^2-m_e^2}{2\sqrt{Q'^2}}$,
$|\bar{p'_1}| = \frac{\lambda^{1/2}(Q'^2,t,m_e^2)}{2\sqrt{Q'^2}}$,
$|\bar{Q}| = \frac{\lambda^{1/2}(Q'^2,Q^2,m_e^2)}{2\sqrt{Q'^2}}$ and
 $\lambda(a,b,c)=a^2+b^2+c^2-2(ab+ac+bc)$. As we chose here
 the z-axis along the $\bar{p'_1}$  (the $\bar p_1$ in the $Q'$ rest frame)
 the vectors have to be rotated
 after generation to restore the general choice of the z-axis
 along $\bar p_1$ in the $e^+e^-$ cms frame.

 Finally the angles of the positively charged pion are generated
 in the $Q$ rest frame with flat distributions

 \bea
\cos\theta_{\pi_1} = -1+ 2r_{\theta_{\pi_1}} \ , \ 
 \phi_{\pi_1} = 2\pi r_{\phi_{\pi_1}} \ .
\eea

The described change of variables
 transforms the phase space into a unit hypercube 
 ($0<r_i<1\ , \ i= Q^2,\cdots , \phi_{\pi_1} $) and collecting
 all the jacobians it reads

\begin{eqnarray}
&&\kern-20pt{\rm dLips}_4(p_1+p_2;q_1,q_2,\pi_1,\pi_2) = P(q_1,q_2)
dr_{Q^2}dr_{Q'^2}dr_{\theta_{q_2}}dr_{\phi_{q_2}}dr_{\theta_Q}dr_{\phi_Q}
dr_{\theta_{\pi_1}}dr_{\phi_{\pi_1}}\non\\
\eea
with
\bea
 P(q_1,q_2)&=& 
 \frac{1}{6(4\pi)^5Q'^2m_e^2} \ \lambda^{1/2}(Q'^2,Q^2,m_e^2) \
\lambda^{1/2}(s,Q'^2,m_e^2) \ \sqrt{1-\frac{4m_{\pi}^2}{Q^2}}
 \non \\   
&&
\frac{t^2\ (Q'^2-m_e^2)^2 \ (Q^2-2Q\cdot p_1)^2 }
 {(Q^2-2E'_{1}Q_0)^2-4|\bar{p'_1}|^2|\bar{Q}|^2} \ 
 \frac{\sqrt{Q^2}(\sqrt{s}-2m_e-2m_{\pi})^2}{\sqrt{s}-2m_e-\sqrt{Q^2}}
\non \\ 
&&\left(\frac{1}{(Q^2+2\sqrt{Q^2}m_e)^3}-\frac{1}{(s-2\sqrt{s}m_e)^3}\right)
\ .
\end{eqnarray}

For the $s$-channel generation it is convenient to write the phase space
 in the following form 
 
 \bea
&&\kern-20pt{\rm dLips}_4(p_1+p_2;q_1,q_2,\pi_1,\pi_2) = \non \\
 &&
 {\rm dLips}_2(p_1+p_2;Q,k_1)\frac{dk_1^2}{2\pi}{\rm dLips}_2(k_1;q_1,q_2)
        \frac{dQ^2}{2\pi}{\rm dLips}_2(Q;\pi_1,\pi_2).
\eea

 The two generation channels used here differ only in the generation
 of the electron--positron pair
 invariant mass $k_1^2$ and the change of variables will be
 described simultaneously. The invariant mass $Q^2$ is generated
 with a flat distribution

\begin{equation}
Q^2 = 4m_{\pi}^2 + ((\sqrt{s}-2m_e)^2-4m_{\pi}^2) r_{Q^2}\ .
\end{equation}

Reflecting two leading $k_1^2$ behaviours of the cross section,
 two distinct changes of variables are done in the generation of $k_1^2$:
%
\bea
k_1^2 &=& s\exp(y_I^{1/3})\ ,
 \label{log1}
 \\
y_I &=& \ln^3(4m_e^2/s)
 +\left[\ln^3\left(\left(1-\sqrt{Q^2/s}\right)^2\right)
 -\ln^3(4m_e^2/s)\right] 
r_{k_1^2,I} \non
\eea
\bea
k_1^2 &=& s\left(1-\exp(-y_{II} ) \right)\ ,
 \label{log2}
 \\
y_{II} &=& - \ln(1-4m_e^2/s) 
 -\ln\left(\frac{\sqrt{Q^2}\left(2\sqrt{s}-\sqrt{Q^2}\right)}
 {(s-4m_e^2)}\right) r_{k_1^2,II} \ . \non
\eea

 The $\bar k_1$ polar angle is used to absorb peaks coming
 from the electron propagator, while its azimuthal angle
 is generated with a flat distribution:

\bea
\phi_{k_1} &=& 2\pi r_{\phi_{k_1}} \ , \ 
 \cos\theta_{k_1} = \frac{-k_1^2+2k_{10}p_{10}}{2|\bar{k_1}||\bar{p_1}|}
\tanh\left(\frac{y}{2}\right)\non \\
 &&\kern-40pt y = \ln \left( \frac
{ k_1^2-2k_{10}p_{10} +{2|\bar{k_1}||\bar{p_1}|}}
{ k_1^2-2k_{10}p_{10} -{2|\bar{k_1}||\bar{p_1}|}} \right)
+ \ln \left( \frac
  { k_1^2-2k_{10}p_{10} -{2|\bar{k_1}||\bar{p_1}|}}
{ k_1^2-2k_{10}p_{10} +{2|\bar{k_1}||\bar{p_1}|}}\right)^2
r_{\theta_{k_1}}  ,
\eea
where $k_{10}= \frac{s+k_1^2-Q^2}{2\sqrt{s}}$,
$p_{10} = \frac{\sqrt{s}}{2}  $,
$|\bar{p_1}| = \sqrt{\frac{s}{4}-m_e^2}$ and
$|\bar{k_1}| = \frac{\lambda^{1/2}(s,k_1^2,Q^2)}{2\sqrt{s}}$, 
are defined in the $p_1+p_2$ rest frame.

 The $\bar q_1$ and the $\bar\pi_1$ angles are generated with flat 
 distributions 

\bea
\kern-10pt
 \phi_{q_1} = 2\pi r_{\phi_{q_1}}, \cos\theta_{q_1} = -1+ 2r_{\theta_{q_1}},
 \phi_{\pi_1} = 2\pi r_{\phi_{\pi_1}},
 \cos\theta_{\pi_1} = -1+ 2r_{\theta_{\pi_1}}\ .
\eea

After the described changes of variables are performed, the phase
 space reads ($i=I\ {\rm or} \ II$) 

\bea
\kern-25pt
{\rm dLips}_4(p_1+p_2;q_1,q_2,\pi_1,\pi_2) =  P_{s,i}
 dr_{k_1^2}dr_{Q^2}dr_{\theta_{k_1}}dr_{\phi_{k_1}}
dr_{\theta_{q_1}}dr_{\phi_{q_1}}dr_{\theta_{\pi_1}}dr_{\phi_{\pi_1}}\ ,
\eea
with 
\bea
\kern-15pt
 P_{s,i} &=&\frac{1}{4(4\pi)^5s} \sqrt{1-\frac{4m_{\pi}^2}{Q^2}}
\sqrt{1-\frac{4m_e^2}{k_1^2}}\lambda^{1/2}(s,Q^2,k_1^2) 
 \left((\sqrt{s}-2m_e)^2-4m_{\pi}^2\right) \non \\  && \cdot
\frac{|\bar{k_1}||\bar{p_1}|}{2k_{10}p_{10}-k_1^2}
\left(\frac{-k_1^2+2k_{10}p_{10}}{2|\bar{k_1}||\bar{p_1}|}
 -\cos\theta_{k_1}\right)
\left(\frac{-k_1^2+2k_{10}p_{10}}{2|\bar{k_1}||\bar{p_1}|}
 +\cos\theta_{k_1}\right) \non
 \\  && \cdot
 P_i  \cdot \ln\left(\frac{-k_1^2+2k_{10}p_{10}+2|\bar{k_1}||\bar{p_1}|}
{-k_1^2+2k_{10}p_{10}-2|\bar{k_1}||\bar{p_1}|}\right)^2 \ ,
\eea
where 
\bea
\kern-25pt
 P_I = \ln^3\left(\left(1-\sqrt{Q^2/s}\right)^2\right)-\ln^3({4m_e^2/s})
 \ , \ 
  P_{II} =\ln\left(\frac{(s-4m_e^2)}{\sqrt{Q^2}(2\sqrt{s}-\sqrt{Q^2})}\right)
\eea
for the change of variables from Eq.(\ref{log1}) or Eq.(\ref{log2})
 respectively. Again $0<r_i<1$ for $i=k_1^2,\cdots,{\phi_{\pi_1}}$. 

 The function, which approximates the peaking behaviour of the
 matrix element reads

 \bea
 F= \left(\frac{1}{P(q_1,q_2)}+\frac{1}{P(q_2,q_1)}
  +\frac{a}{P_s}\right)^{-1} \ , \ {\rm with} \ 
P_s = \frac{P_{s,I}+\mathrm{b}P_{s,II}}
{\frac{3\ln^2(k_1^2/s)}{k_1^2}+\frac{\mathrm{b}}{s-k_1^2}} \ .
 \eea 

The introduced a priori weights $a$ and $b$, which guarantee the right
 relative contributions from different generation channels,
  were set to $a=1.1$ and $b=1000$, a choice optimal for DAPHNE energy.

\section{The pair production and the radiative return}
\label{sec1}
 
\begin{figure}[ht]
\begin{center}
\epsfig{file=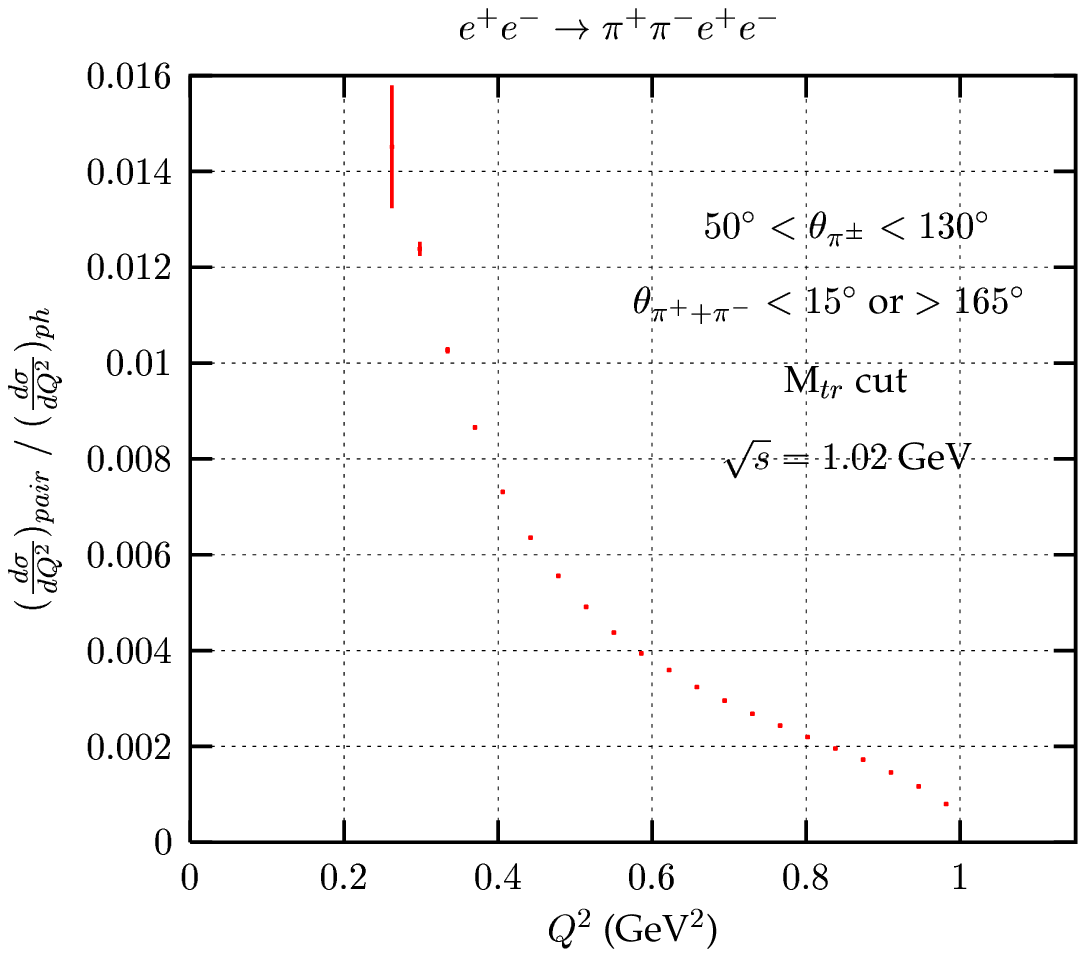,width=6.5cm,height=6.3cm}
\hskip+0.5cm
\epsfig{file=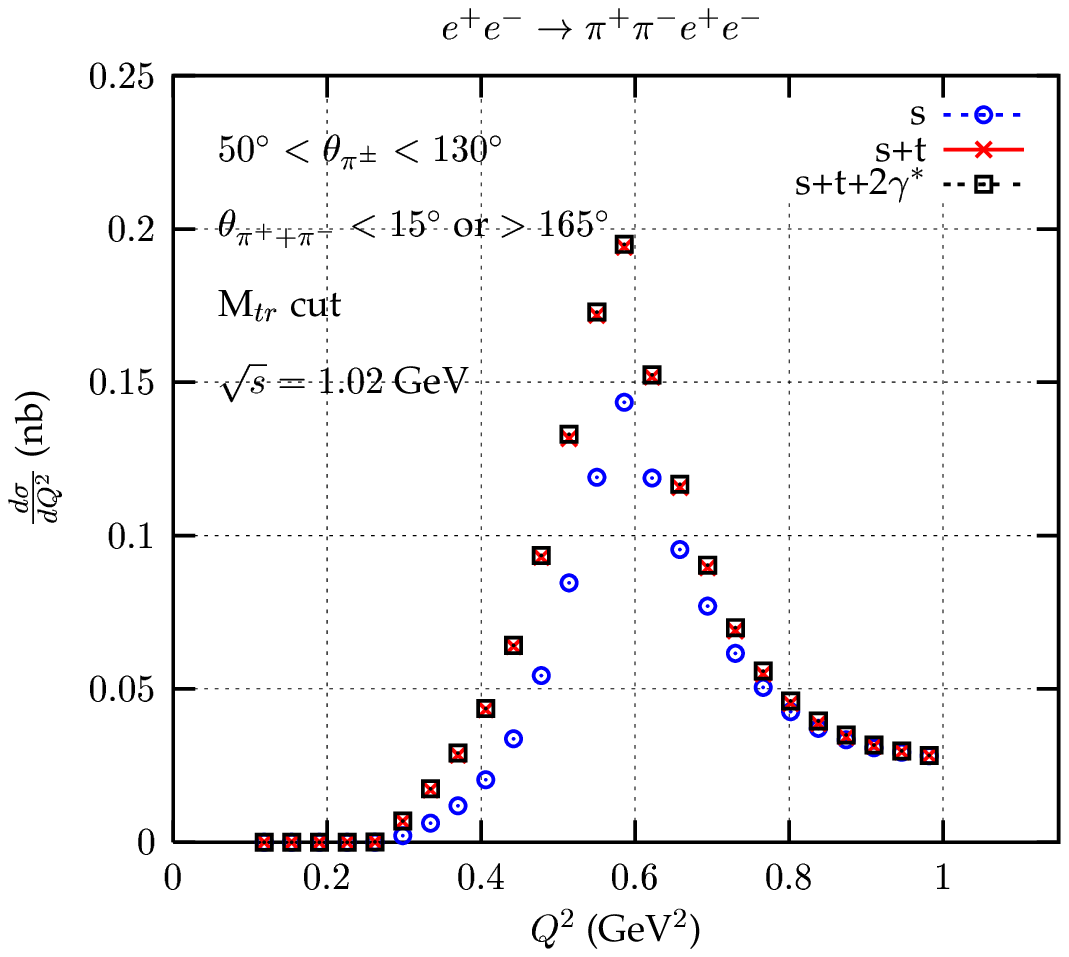,width=6.5cm,height=6.3cm}
\hskip+0.2cm
\end{center}
\caption{The ratio of the differential cross sections of
 the reactions $e^+e^-\to e^+e^-\pi^+\pi^-$ and
  $e^+e^-\to\pi^+\pi^-\gamma(\gamma)$ for KLOE \cite{KLOE1}
  event selection (left) and separate contributions to the cross
 section of the reaction $e^+e^-\to e^+e^-\pi^+\pi^-$:
 $s$ -- diagrams from Fig. \ref{diag}a-\ref{diag}c included;
 $s+t$ -- diagrams from Fig. \ref{diag}a-\ref{diag}d included;
 $s+t+2\gamma^*$ -- all diagrams from  Fig. \ref{diag} included (right).
}
\label{KLOE}
\end{figure}

 As the only experiment, which uses the radiative return method
 without photon tagging is KLOE, we present here the results
 for DAPHNE energy only.
 The cross section of the reaction  $e^+e^-\to\pi^+\pi^-\ + \ \gamma(\gamma)$
 was obtained with the PHOKHARA~5.0 \cite{Czyz:2005as} event generator.
 For the event selection used by KLOE
 \cite{KLOE1} the relative contribution of the pair production
 to the cross section of the reaction 
 $e^+e^-\to\pi^+\pi^-\ + \ \gamma(\gamma)$
 is shown in Fig.\ref{KLOE} (left). It amounts up to 1.5\% in
 the vicinity of the production threshold, but it is below 1\%
 in the $Q^2$ region, where the measurement was performed
 \cite{KLOE1} ($Q^2>~0.33 {\rm GeV}^2$).

 \begin{figure}[ht]
\begin{center}
\epsfig{file=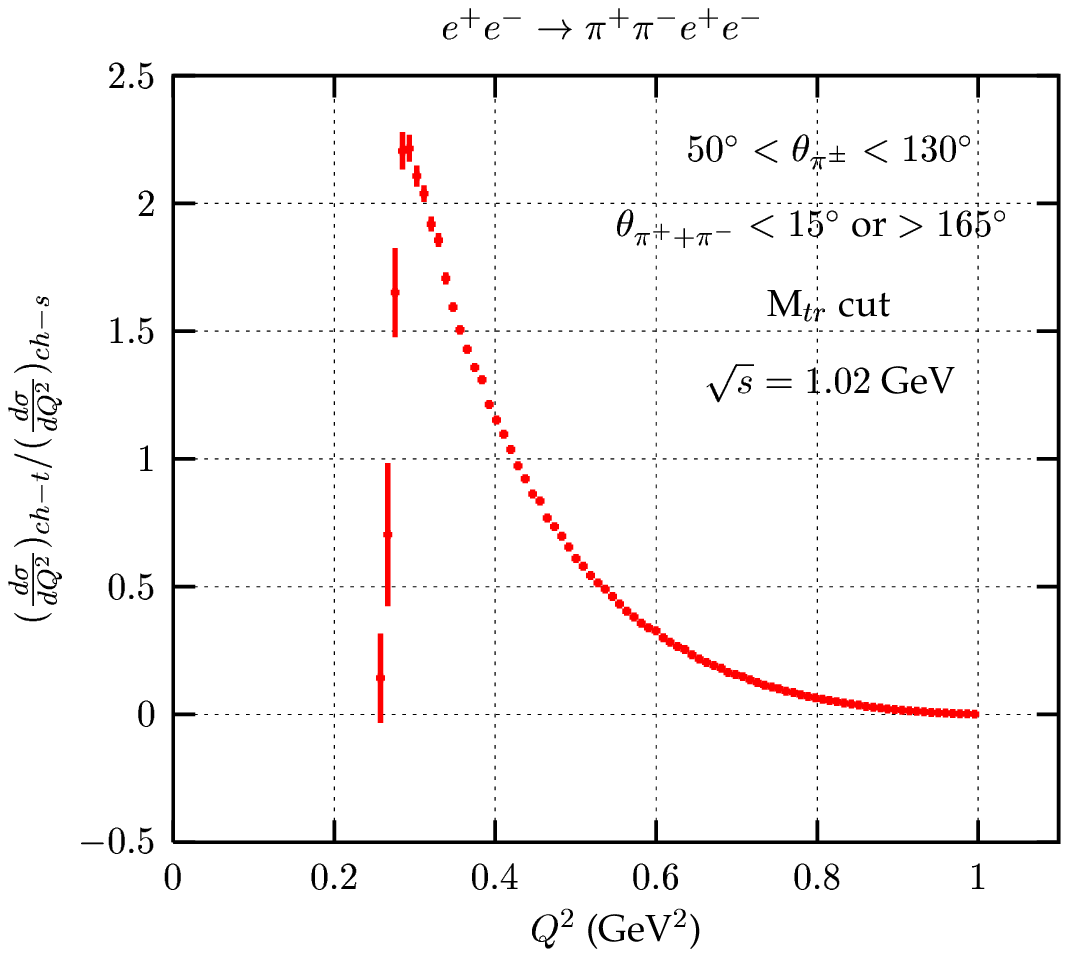,width=6.5cm,height=6.3cm}
\hskip+0.5cm
\epsfig{file=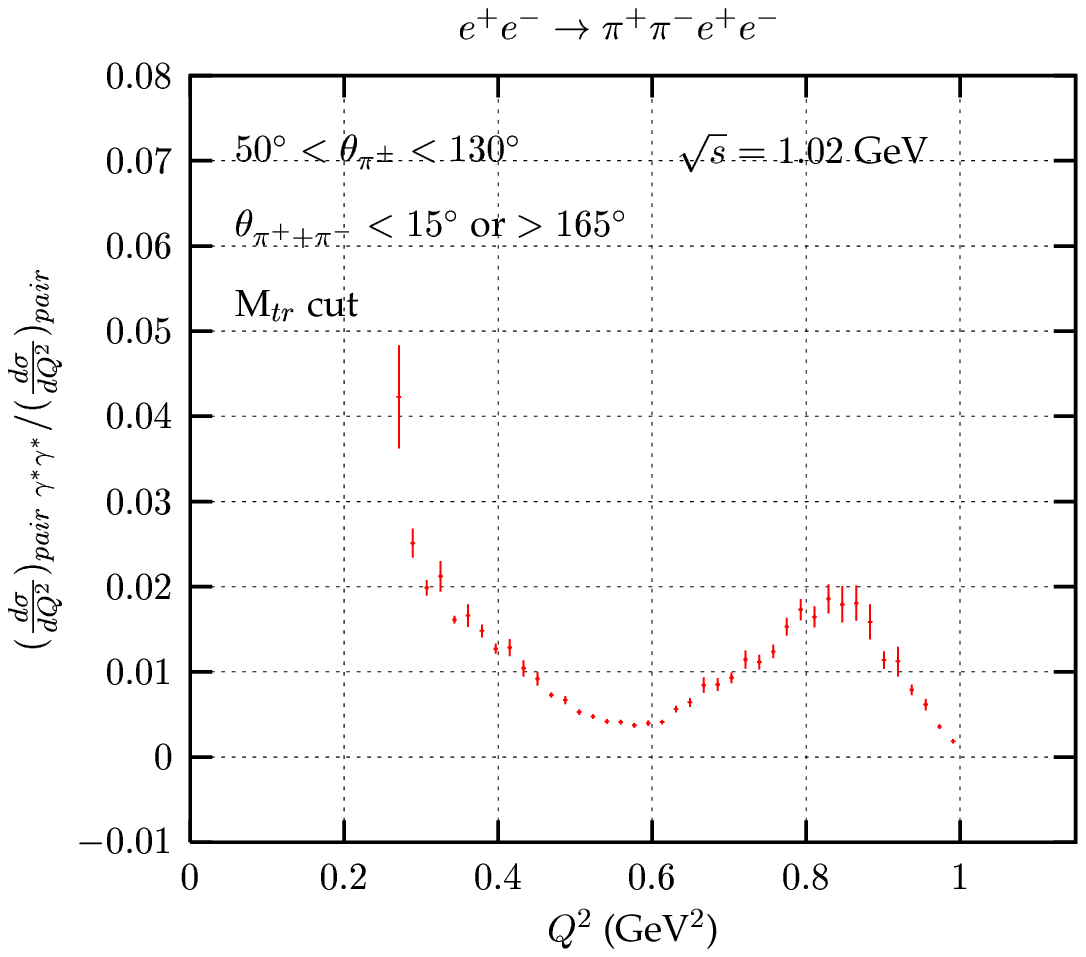,width=6.5cm,height=6.3cm}
\hskip+0.2cm
\end{center}
\caption{The ratio of the t-channel and the s-channel contributions
to the differential cross section 
of  the  reaction $ e^+e^- \to \pi^+\pi^-e^+e^- $  (left);
 the relative contribution of the photon fusion process to
 the differential cross section  
 of  the  reaction $ e^+e^- \to \pi^+\pi^-e^+e^- $  (right)}
\label{KLOE1}
\end{figure}

\begin{figure}[ht]
\begin{center}
\epsfig{file=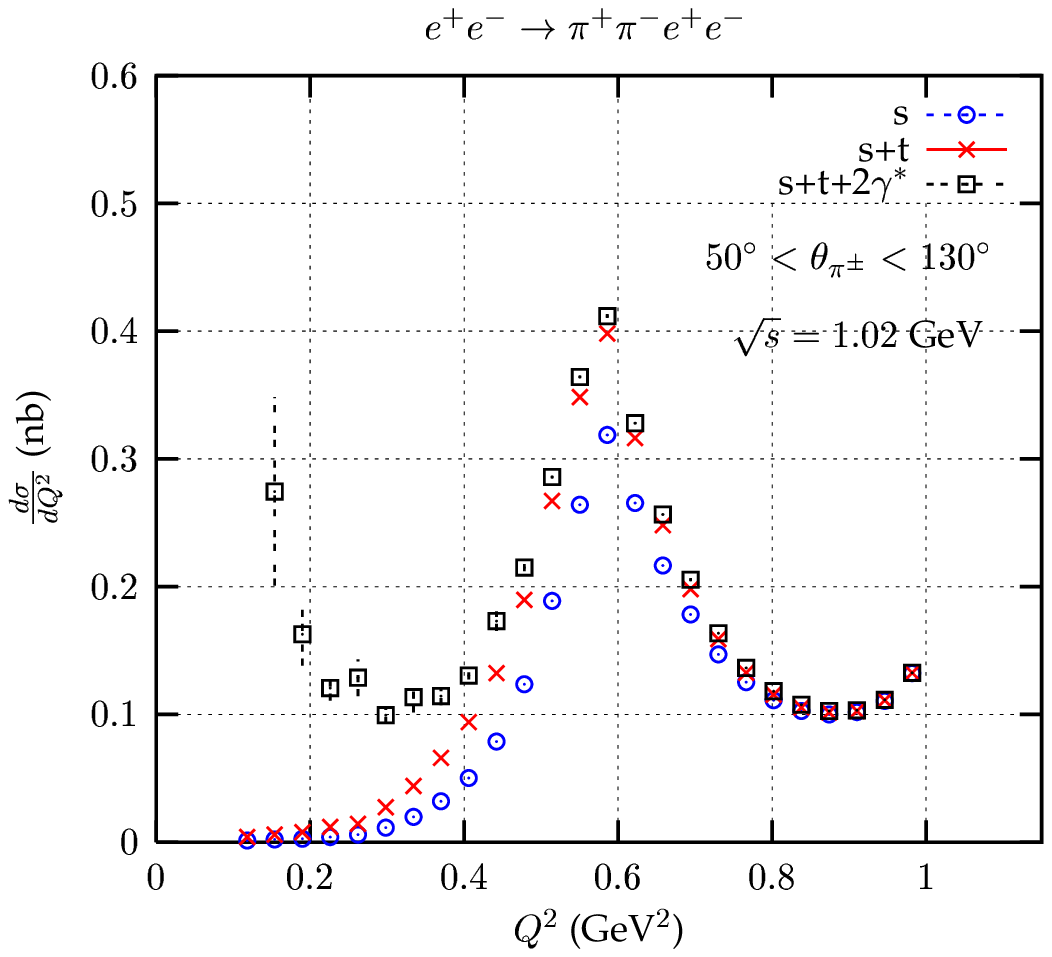,width=6.4cm,height=6.3cm}
\epsfig{file=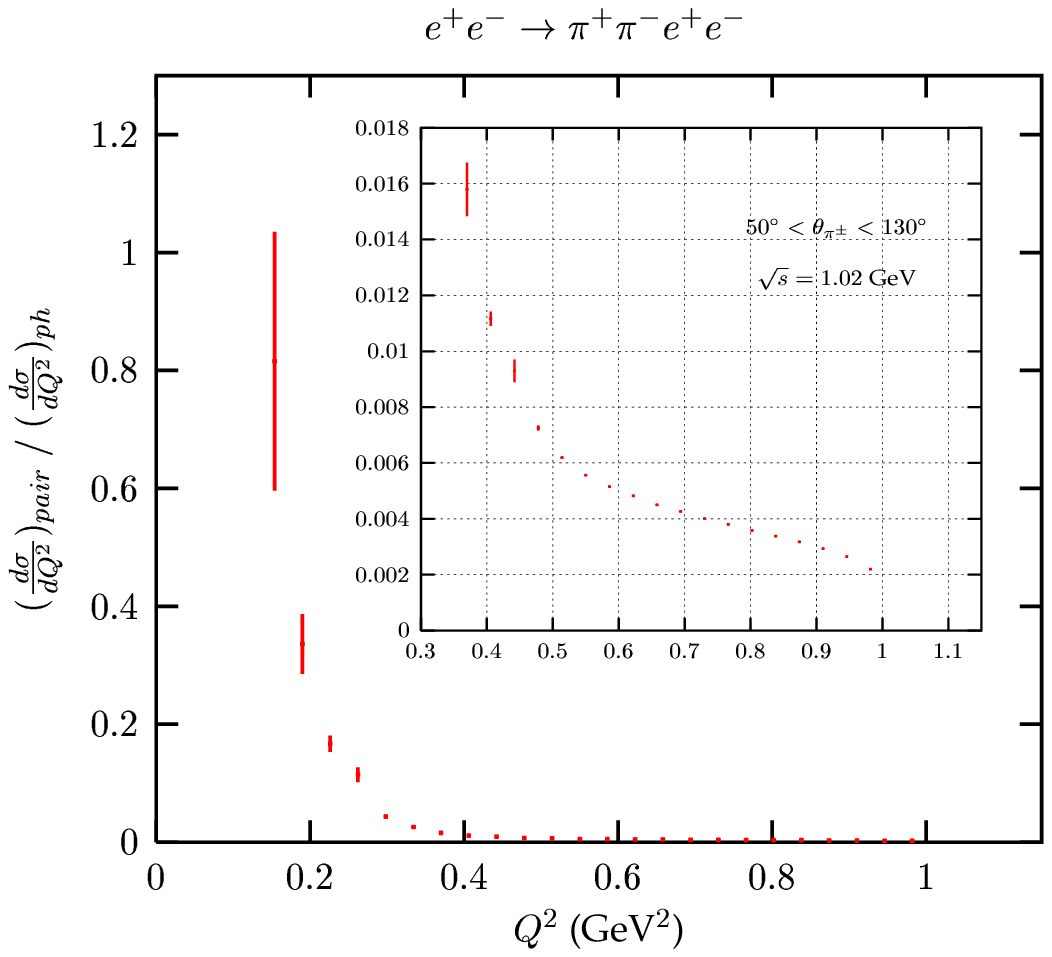,width=6.4cm,height=6.3cm}
\end{center}
\caption{Various contributions to differential cross sections of the 
 reaction $e^+e^-\to e^+e^-\pi^+\pi^-$: 
 $s$ -- diagrams from Fig. \ref{diag}a-\ref{diag}c included;
 $s+t$ -- diagrams from Fig. \ref{diag}a-\ref{diag}d included;
 $s+t+2\gamma^*$ -- all diagrams from  Fig. \ref{diag} included (left);
  The ratio of the differential cross sections of
 the reactions $e^+e^-\to e^+e^-\pi^+\pi^-$ and
  $e^+e^-\to\pi^+\pi^-\gamma(\gamma)$ (right).
 In both cases only cuts on pions polar angles are imposed.}
\label{angular}
\end{figure}

 For small $Q^2$ values the main 
 contribution is given by the $t-$channel diagrams, while around
 the $\rho$ resonance the s-channel dominates 
 (see Fig.\ref{KLOE}(right) and Fig.\ref{KLOE1}). In the latter case,
  as it was shown in
 \cite{n1}, the ISR contributions dominate and the FSR is irrelevant.
 Moreover, the contributions  to the cross sections
 from diagrams from Fig.\ref{diag}c are always small and 
 the two-photon contributions (Fig.\ref{diag}e) are completely negligible
 for the KLOE event selection as already pointed out in \cite{Juliet}
 and shown in  Fig.\ref{KLOE1}(right). This may not be true for different event
 selections, as the photon fusion diagrams can give sizable contributions
 especially in low $Q^2$ region. An example is presented in Fig.\ref{angular},
 where only cuts on pion angles are applied. The pair production cross
 section can be in this case
 almost as big as the $e^+e^-\to\pi^+\pi^-\gamma(\gamma)$
 cross section, where the $s+t$ channel diagrams (Fig.\ref{diag}(a-d))
 give only up to 3\% of the
 $e^+e^-\to\pi^+\pi^-\gamma(\gamma)$ cross section \cite{n2} and
 the main contribution is given by the photon fusion diagrams.

 In the case of s-channel ISR and the pion pair emission from
 t-channel Bhabha diagrams a factorization analogous to the one of 
 Eq.(\ref{radiator})
 occurs, with the radiator function given by QED. As a result, for the case
 of the KLOE event selection without photon tagging, it is possible to use the 
 radiative return method adding contributions from the photon(s) and
 the pair production. The pion form factor, 
 to be extracted from the data, is the same in both cases and the radiator
 function is a sum of both contributions.
 This procedure will be necessary, when the accuracy of
 the measurement will be  below 1\% at low $Q^2$ values.
 Alternatively, one can treat the pair production as a backgroud
 and study carefully the accuracy of its estimation.

\section{Conclusions}
\label{sec2}
 Basing on the developed Monte Carlo program EKHARA presented in this letter,
 it was shown that 
 the reaction $e^+e^-\to e^+e^-\pi^+\pi^-$
 may give non negligible contributions to the pion form factor measurement
  via the radiative return method without photon tagging.
 For low invariant masses of the two--pion system it is up to 1\%
 for the event selection used in KLOE analysis \cite{KLOE1},
 but it can be substantially larger if some of the cuts, used
 in the analysis, are relaxed.

\section{Acknowledgments}

We would like to thank Achim Denig, Wolfgang Kluge, Debora Leone 
 and Stefan M\"uller for 
discussions of the experimental aspects of our analysis and TTP of
 the Univ. of Karlsruhe, where a part of this work was performed,
 for the kind hospitality.

\end{document}